\documentstyle[aps,prl,epsf]{revtex}
\begin{document}
\draft
\twocolumn[\hsize\textwidth\columnwidth\hsize\csname %
@twocolumnfalse\endcsname

\title{Double exchange magnets: Spin-dynamics in the paramagnetic phase} 

\author{Peter Horsch$^a$, Janez Jakli\v c$^{a,b}$, and Frank Mack$^a$}

\address{
$(a)$ Max-Planck-Institut f\"{u}r Festk\"{o}rperforschung,
Heisenbergstr.~1, D-70569 Stuttgart (Germany) \\
$(b)$ Max-Planck-Institut f\"{u}r Physik komplexer Systeme, Bayreuther Str.~40,
D-01187 Dresden (Germany)}

\date{\today}
\maketitle

\begin{abstract}
The electronic structure of perovskite manganese oxides is
investigated in terms of a Kondo lattice model with ferromagnetic Hund
coupling and antiferromagnetic exchange between $t_{2g}$-spins using a
finite temperature diagonalization technique.
Results for the dynamic structure factor
are consistent with recent neutron scattering experiments
for the bilayer manganite La$_{1.2}$Sr$_{1.8}$Mn$_2$O$_7$ .
The susceptibility shows Curie-Weiss behaviour and is used
to derive a phase diagram.
In the paramagnetic phase carriers are characterized as ferromagnetic 
polarons in an antiferromagnetic spin liquid.
\end{abstract}
]
\narrowtext

Double exchange (DE) as an explanation of ferromagnetic order in doped
manganites\cite{Zener51} 
was proposed shortly after the first experimental work on
these compounds\cite{Jonker50}. 
Open
questions are connected with the proper understanding of the
paramagnetic (PM) phase above the Curie temperature $T_C$ and the giant
magnetoresistance at the phase transition.
Millis {\it et al}\cite{Millis95a} argued that DE is neither
sufficient to explain the size of the resistivity above $T_C$ nor the
activated form of $\rho(T)$ observed in many compounds, and suggested 
that Jahn-Teller polaron formation is essential.
A different point of view \cite{Varma96}, 
is that these
systems are governed by strong correlations while spin disorder
scattering is crucial for the understanding of the resistivity.
Further complications not considered here
may arise from orbital degeneracy\cite{Kugel73,Inoue97} 
and charge disorder\cite{Allub97}. 
 
A detailed investigation of the spin-dynamics
is therefore important for a deeper understanding 
of the manganites. Inelastic neutron scattering
experiments in the PM-phase were performed recently by  
Perring {\it et al.}\cite{Perring97} for the highly doped bilayer manganite system 
La$_{1.2}$Sr$_{1.8}$Mn$_2$O$_7$\cite{Moritomo96}. 
Particularly surprising was the observation of growing
antiferromagnetic (AF) scattering at low energy parallel to increasing
ferromagnetic (FM) critical scattering as the FM $T_C$
was approached from above.
Our aim here is to shed light on
these experiments by an investigation of the spin-dynamics within
a standard model for these compounds.

The appearance of ferromagnetism was explained by Zener\cite{Zener51}
in terms of a strong Hund coupling $J_H$ of the $e_g$
conduction electrons to S=3/2 core spins ${\bf S}^c_{\bf i}$ 
formed by the energetically well separated  $t_{2g}$-electrons.
The $e_g$ electron in Mn$^{3+}$ couples to a S=2 high-spin
configuration. A relevant model is the {\it ferromagnetic} Kondo
lattice model\cite{Kubo72,Riera96,Kaplan97}
\begin{eqnarray}
H_{KLM} =&-&\sum_{{\bf ij}\sigma} t_{{\bf ij}}
d_{{\bf i}\sigma}^{\dagger} d_{{\bf j}\sigma}
-J_H\sum_{{\bf i}\sigma\sigma'} {\bf S}^c_{{\bf i}}\cdot
d_{{\bf i}\sigma}^{\dagger}\vec {\sigma}_{\sigma\sigma'}d_{{\bf i}\sigma'}\nonumber\\
&+& J_{AF}\sum_{<{\bf i j}>} {\bf S}^c_{{\bf i}}\cdot {\bf S}^c_{{\bf j}}
+ U\sum_{{\bf i}} n_{{\bf i}\uparrow} n_{{\bf i}\downarrow} .
\label{eq:FKLM}
\end{eqnarray}
The operators $d_{{\bf j}\sigma}$ and ${\bf S}^c_{{\bf i}}$ refer to
$e_g$-electrons and $t_{2g}$-spins, respectively, while 
$n_{{\bf i}\sigma}=d_{{\bf i}\sigma}^{\dagger} d_{{\bf i}\sigma}$
denotes the $e_g$-density operator.
A typical value for the Hund coupling is $J_H\sim 1$eV, while 
the nearest neighbor hopping $t_{\bf ij}=t\sim
0.15$eV\cite{Inoue97,Satpathy96}. 
In view of the large local repulsion $U \sim 8$ eV, we shall
furtheron exclude double occupancy of $e_g$-orbitals, 
and define $J_H/t=6$.
The model explains ferromagnetism via the optimization of the kinetic
energy of the $N_e=\sum n_{{\bf i} \sigma}$
conduction electrons in the partially filled $e_g$ band
when the spins are aligned at low temperature. 
At low doping manganite materials are usually AF
insulators. Antiferromagnetic super-exchange 
interaction $J_{AF}$ between the $t_{2g}$-electrons or spins is an
obvious source of such correlations.

The interplay of the global AF exchange interaction
with the local FM alignment
induced by the motion of $e_g$-electrons via the DE mechanism 
is a subtle problem. 
A simple physical picture, which is consistent with our numerical
study,  is that of {\it ferromagnetic} spin
polarons, where a local FM surrounding of $t_{2g}$-spins moves
with the holes through an otherwise AF background.

Our calculations are based on a finite temperature 
diagonalization approach developed by Jakli\v c and 
Prelov\v sek\cite{Jaklic94}, which was
successfully used for the study of the $t$-$J$ model. 
Here due to the much larger Hilbert space only relatively small 
systems can be solved. Nevertheless due to the high density of low
energy excitations we believe that our results reflect already the
physics of large systems. We have found large similarity between 1D
and 2D systems, in striking contrast to the physics of the $t$-$J$
model. We shall present here mainly 1D results,
since the resolution in momentum space is larger.
Our energy and temperature units are $t=1$, while $S=1$ is used
instead of $S=3/2$ $t_{2g}$-spins to save computer time.

A global measure for the evolution of
magnetic correlations is given by the square
of the total spin ${\bf S}=\sum {\bf S}_{\bf i}$, which includes both
core- and $e_g$-spins.
Results for different $e_g$-concentration and $J_{AF}$ are shown in
Fig. 1(a). The high-$T$ limit is determined by independent spins
(i.e. for $T < J_H$; for even higher $T$ also the $e_g$-spins decouple).       
In the absence of  $J_{AF}$ FM
correlations develop as the temperature is lowered.
For quite small $J_{AF}$ this behaviour is already changed. 
For $J_{AF}=0.1$  $\langle {\bf S}^2\rangle$ decreases with decreasing
temperature for a single hole. 
This suggests the evolution of AF
correlations, while at higher doping FM
correlations are established via the DE mechanism.
Further insight is gained from the uniform susceptiblity
$\chi=\langle S^2_z\rangle /Nk_BT$.
The corresponding $1/\chi(T)$-data, Fig. 1(b), show Curie-Weiss
behavior,
similar to experiments\cite{Donnell96,Oseroff96}. 
It is quite remarkable that for these small systems with $N=6$ and $8$
sites $1/\chi(T)$
follows a linear Curie-Weiss law to quite low temperatures, which
allows us to determine a mean-field transition temperature $T_C^{MF}$. 
Deviations at lower temperatures are expected from non mean-field
behavior of low-dimensional systems and finite size effects.
\\
\\
\\
\begin{figure} 
\epsfxsize=6cm
\epsffile[60 20 530 680]{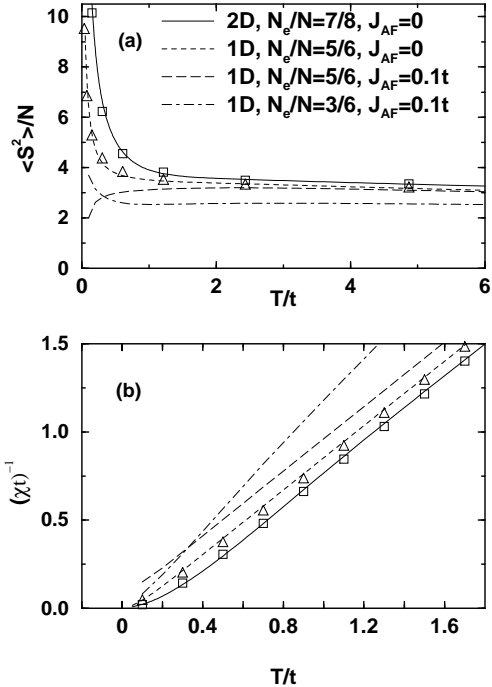}
\caption{ Square of total spin (a) and inverse
  susceptibility (b) vs. temperature  for hole concentrations 
  $x=(1-N_e/N)=1/6, 3/6$ (1D) and $1/8$ (2D). 
  A comparison with the FM polaron
  approach is shown for a single hole 
  in 1D (triangles) and 2D (squares) for $J_{AF}=0$.
\label{fig1}}
\end{figure}

The mean-field phase diagram (Fig. 2)
reflects the interplay between antiferromagnetism
and carrier induced ferromagnetism at higher doping. 
$T_C^{MF}$ scales
with coordination number and is consistent with the experimental
$T_C$-scale. 
The value chosen for $J_{AF} (\sim 0.1)$ is typical for real systems
\cite{Perring97,Inoue97}.
For $J_{AF}=0$ our data is well described by ${T_C^{MF}} \sim x(1-x)$ 
\cite{Varma96}, while it differs in that respect from high-temperature
series expansion results\cite{Roeder97} .

A simple estimate of  $\langle {\bf S}^2\rangle$ 
in terms of a {\it ferromagnetic}
spin-polaron picture has been given by Varma\cite{Varma96}
for the model without AF-interactions. The
spin polaron is assumed to be determined by a number $P$ of
ferromagnetically aligned spins around the hole which allows the
particle to improve its kinetic energy. The gain in kinetic energy
is counterbalanced by the loss of spin entropy due to the coupling of
P spins, which can no longer rotate independently.  An estimate for
the change of free energy is
$\delta E_P = a t/P^{\eta} + P k_B T \ln(2 S_2 +1)$.
The kinetic energy exponent $\eta=2/3, 1, 2$ in 3, 2, and
1-dimensions, respectively. Minimization leads to 
$P(T)=(\frac{\eta a t}{B k_B T})^{1/(1+\eta)}$ with $B=\ln(2 S_2 +1)$.
Under the same assumptions the square of the total spin  
is given in terms of $S_P=(S_1+P S_2)$
\begin{equation}
\langle{\bf S}^2\rangle=
N\Bigl[ x S_P (S_P +1)+(1-x-P x) S_2 (S_2+1)\Bigr],
\end{equation}
where in the case of few $e_g$-holes $x$ is the 
hole  concentration. Here $S_1=1$ and $S_2=3/2$ (instead of 3/2 and 2).
A comparison of this approach with exact diagonalization is shown in
Figs. 1(a) and (b) for a 1D (2D) system with 6(8)-sites and one $e_g$
hole. 
Good fits over the whole $T$-range were obtained by choosing smaller
values for the exponent $\eta = 1.5 (0.6)$ for 1D (2D), respectively. 
 
The effective spin $S_{eff}$ for a single polaron may be defined as
$S_{eff}(S_{eff}+1)=\langle {\bf S}^2 {\rangle}_T
-\langle {\bf S}^2 {\rangle}_{\infty}$. Near $T_C^{MF}$ we
obtain $S_{eff}\sim 7$ for the two-dimensional case. This corresponds
to an enhancement factor $\sim 14$ with respect to the spin $S=1/2$ of
the $e_g$ conduction electron (hole).
This result is qualitatively consistent with EPR experiments 
\cite{Oseroff96} where large effective spins at temperatures  
slightly above $T_C$ were observed.
\\
\begin{figure} 
\epsfxsize=5cm
\epsffile[60 20 430 420]{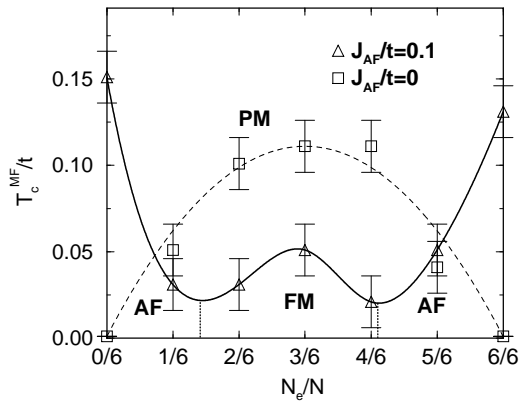}
\caption{Mean-field phase diagram of a 1D chain for
  $J_{AF}=0.1$ (triangles) vs. $(1-x)=N_e/N$.
  The boundaries between AF, FM and PM phases (solid
  lines) are guides to the eye. 
  (Anti)periodic boundary conditions were used for odd (even) electron
  number [18].
  The $J_{AF}=0$ data (squares) is compared with the predicted
  behavior $T_C^{MF} \sim x(1-x)$.
\label{fig2}}
\end{figure}

The study of correlation functions (CF)
gives further support to the FM polaron picture even at
high doping concentration and with frustrating AF
interactions, although the simple formulation above is
certainly not sufficient for these cases. 
The temperature dependence of the $ \langle {\bf S}_i\cdot {\bf S}_j \rangle$
correlation functions for $J_{AF}=0.1$ is shown in Fig. 3.
We distinguish (a) a low- and (b) a high-doping case where the system
approaches the AF (FM) phase (Fig. 2) at low
temperature, respectively.
In the single hole case (a) the nearest-neighbor (n.n.) CF becomes 
AF as T is
lowered, while further neighbor correlations remain small. The CF's
for $x=0.5$ (b) become all FM at low T. Yet it is
interesting to notice that for intermediate T the n.n. CF is 
antiferromagnetic. For $T \gg t$ all CF's vanish.

To investigate the structure of the spin-polaron we calculate
the {\it relative correlation function} (RCF)  
\begin{equation}
C_{\bf n}^{\bf n+l} = \sum_{\bf i} \langle n^h_{\bf i} {\bf S}_{{\bf i+n}}\cdot 
{\bf S}_{{\bf i+n+l}} \rangle,
\label{eq:RCF}
\end{equation}
which measures the spin-CF
$  \langle {\bf S}_{{\bf 0}}\cdot {\bf S}_{{\bf l}}  \rangle $
in a distance ${\bf n}$ from the position of the moving hole. Here
$n^h_{\bf i}= (1-n_{\bf i})$ and ${\bf l}$ is a lattice vector. 
The results in Fig. 3 show the FM
alignement of the spins at the position of the hole and on the
neighboring site, i.e.  $C_{0}^{1}$, at low T even in the single hole
case where $  \langle {\bf S}_{{\bf 0}}\cdot {\bf S}_{{\bf 1}}
\rangle $ becomes antiferromagnetic. 
The large drop at very small $T$ suggests that FM-polarons are
unstable in the low temperature AF-phase consistent with Fig. 1(a).
The CF's $C_{0}^{1}$ are weakly antiferromagnetic at higher temperatures
similar to $ \langle {\bf S}_i\cdot {\bf S}_j \rangle$.

\begin{figure} 
\epsfxsize=6cm
\epsffile[60 20 530 680]{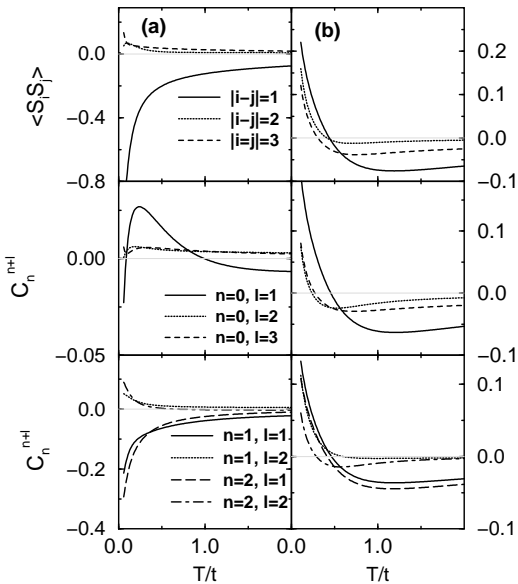}
\caption{Temperature dependence of spin correlation functions
   $ \langle {\bf S_iS_j} \rangle$ and $C_{\bf n}^{\bf n+l}$ (Eq. 3) for one (a) and 3
   $e_g$-holes (b) for a 6-site chain with $J_{AF}=0.1$.
\label{fig3}}
\end{figure}
In the single hole case (a)
all other ($n\neq 0$) RCF's become gradually AF with
decreasing temperature, i.e. indicating that the AF-phase dominates.
The CF's for 
$x=0.5$ (b) are indicative for frustrated antiferromagnetism at 
temperatures above $T \sim 0.5$, while at lower temperature all CF's
become ferromagnetic.
Hence on the basis of these {\it equal-time} CF's there is
no evidence for
significant AF correlations at low $T$ near $T_C^{MF}$ for
high doping. 
Nevertheless there are clear AF-correlations in the low-frequency
response.

The complete information about the spin-dynamics is contained in the
dynamic structure factor
\begin{equation}
S_{\bf q}(\omega)=\frac{1}{2\pi}\int^{\infty}_{-\infty} dt
\exp{(-i\omega t)}\langle {\bf S}_{\bf q}{\bf S}_{-\bf q}(t)\rangle ,
\label{eq:Sofqw}
\end{equation}
which can be directly measured by inelastic neutron scattering.
A central result of this work is the evolution of $S_{\bf q}(\omega)$
with temperature (Fig. 4). As an example results for a 6-site chain
with 3 $e_g$-holes are shown for $J_{AF}=0$ (a) and $0.1$ (b) at
different momenta.
At high $T$ critical scattering extends over the whole Brillouin zone,
while at low temperature
it is confined to $q\sim 0$ in the absence of 
AF-interactions. In case (a) FM
magnons emerge when lowering the temperature already well above
$T_C^{MF}\sim 0.1$. 
Their energy scale is consistent with theoretical work on the 
magnon dispersion\cite{Furukawa96} and
neutron scattering in the FM-phase\cite{Perring96} of
La$_{0.7}$Pb$_{0.3}$MnO$_3$. Antiferromagnetic frustration appears to be
small in this system, since already LaMnO$_3$ consists of FM planes with a
well developed magnon spectrum even at $T_C$\cite{Hirota96}.
The interlayer coupling is AF and weak in this system.
\\
\\
\\
\begin{figure} 
\epsfxsize=6cm
\epsffile[80 80 530 680]{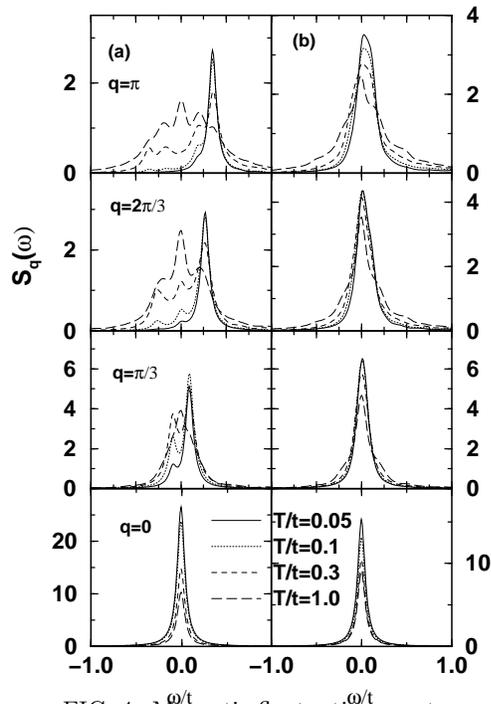}
\caption{Magnetic fluctuation spectrum  $S({\bf q},\omega)$ as function
   of $\omega$ for different momenta and temperatures. Data for a
   6-site chain with $e_g$-electron concentration $x=0.5$ for 
   (a) $J_{AF}=0$ and (b) $J_{AF}=0.1$.
\label{fig4}}
\end{figure}
In the frustrated case (b) the
magnon structures are hardly visible, consistent with a substantial softening.
The dominant feature is
{\it critical scattering for all} $q$. 
As temperature is lowered FM-critical scattering is dominant,
yet surprisingly
low energy AF-correlations at $q=\pi$ are growing as well.
We note that our data is broadened by $\delta=0.05$. 
The natural linewidth of the $\omega\sim 0$ structures appears to be 
smaller than T.

The low frequency response 
$ \delta S({\bf q})=\int^{\omega_c}_{-\omega_c} d\omega S_{\bf q}(\omega)$ 
integrated over a small energy window ($\omega_c=0.05)$, 
is frequently measured in
neutron scattering experiments\cite{Perring97}. 
It is interesting to contrast the momentum dependence of 
$ \delta S({\bf q})$
from the behavior of the static structure factor
$ S({\bf q})=\int^{\infty}_{-\infty} d\omega S_{\bf q}(\omega)$ ,
which measures equal-time correlations (Fig. 5). In the
high-temperature limit $ S({\bf q}) $ is constant, reflecting
the absence of spin-correlations. At low temperature 
$ S({\bf q})$ develops a peak centered at $q=0$ in the model
without antiferromagnetic interactions due to carrier induced
ferromagnetic correlations. 
With increasing $J_{AF}$ the
data shows the evolution of short-range (nearest-neighbor) AF
correlations. 
\\
\\
\begin{figure} 
\epsfxsize=6cm
\epsffile[60 20 530 680]{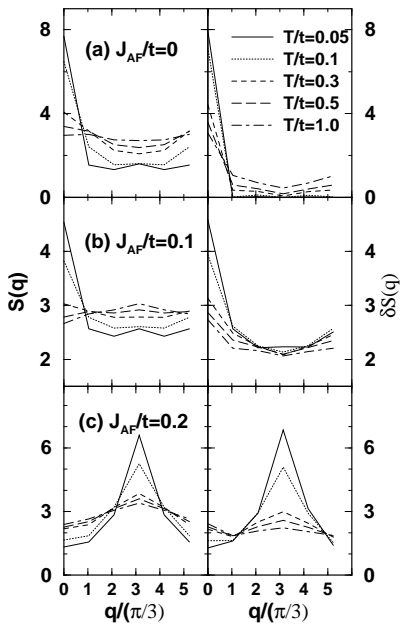}
\caption{Comparison of $S({\bf q})$ and integrated low-frequency response
$\delta S({\bf q})$; 
parameters as in Fig. 4.
\label{fig5}}
\end{figure}
The observation of pronounced AF-correlations in  
$ \delta S({\bf q})$, e.g. for $J_{AF}=0.1$,
in the presence of FM critical scattering is consistent with the
experimental finding of local AF correlations by
Perring {\it et al}\cite{Perring97}. These experiments show strong
AF correlations in the PM-phase 
(at low frequency) and their
sudden disappearence at $T_C$ when the low-$T$ FM-order is established. 
We stress that the observation of a
well defined  AF peak in the
low-frequency spin-fluctuation spectrum $S_{\bf q}(\omega\rightarrow
0)$  is actually very surprising since these
experiments were performed for a highly doped ($x=0.4$) double-layer
system and at temperatures
slightly above the transition into the FM-phase.
We find considerable qualitative agreement between the
present model and neutron scattering, although we do not find a significant
peak structure in the momentum dependence at $q=\pi$, except for
larger values for $J_{AF}$ where the FM scattering is reduced.
The disappearance of the AF signal in the low-temperature
FM phase\cite{Perring97} 
is consistent with our studies\cite{Mack97}.

In summary we have shown, that the competition between
AF and FM correlations detected by neutron
scattering in the highly doped double layer compound
La$_{1.2}$Sr$_{1.8}$Mn$_2$O$_7$ is consistent with the physics of the
frustrated Kondo lattice model.
From our study of various correlation functions, we conclude that the
temperature and doping dependence of the magnetic correlations 
in the {\it paramagnetic phase} can be qualitatively
understood in terms of FM polarons moving in a short-range
AF-spin liquid. 
Slow spin-dynamics as a consequence of frustration may lead to
localization of magnetic polarons.

We would like to thank G. Khaliullin, K. Kubo and A. M. Oles
for stimulating discussions. Partial support by E.U. Grant No. ERBCHRC
CT94.0438 is acknowledged.


\begin{references}
\bibitem{bc}
{Alternating boundary conditions were chosen to favor high-spin
states and to guarantee a systematic dependence on doping.
See also \cite{Riera96,Kaplan97} .}
\end{references}
\end{document}